# An Electrically Conductive Oleogel Paste for Edible Electronics

*Pietro Cataldi\*, Leonardo Lamanna, Claudia Bertei, Federica Arena, Pietro Rossi, Mufeng Liu, Fabio Di Fonzo, Dimitrios G. Papageorgiou, Alessandro Luzio, Mario Caironi\**

Dr. P. Cataldi, Dr. L. Lamanna, F. Arena, P. Rossi, Dr. F. Di Fonzo, Dr. A. Luzio, Dr. M. Caironi
Center for Nano Science and Technology @PoliMi, Istituto Italiano di Tecnologia, Via Giovanni Pascoli, 70/3, Milano, 20133 Italy

C. Bertei, Dr. D. G. Papageorgiou
School of Engineering and Materials Science, Queen Mary University of London, Mile End Road, London E1 4NS, UK

F. Arena
Politecnico di Milano, Department of Energy, Via Lambruschini 4, Milano, Italy

P. Rossi
Politecnico di Milano, Department of Physics, P.zza Leonardo da Vinci 32, 20133, Milano, Italy

Dr. M. Liu
National Graphene Institute, Henry Royce Institute and Department of Materials, University of Manchester, Oxford Road, Manchester M13 9PL, UK

Edible electronics will facilitate point-of-care testing through safe and cheap devices that are digested/degraded in the body/environment after performing a specific function. Its thrive depends on creating a library of materials that are the basic building blocks for eatable technologies. Edible electrical conductors fabricated with green methods that allow production at a large scale and composed of food derivatives, ingestible in large amounts without risk for human health are needed. Here, conductive pastes made with materials with a high tolerable upper intake limit (≥ mg/kg body weight /day) are proposed. Conductive oleogel paste composites, made with biodegradable and food-grade materials like natural waxes, oils, and activated carbon conductive fillers, are presented. The proposed pastes are compatible with manufacturing processes such as direct ink writing and thus are suitable for an industrial scale-up. These conductors are built without the use of solvents, and with tunable electromechanical features and adhesion depending on the composition. They have antibacterial and hydrophobic properties, so that they can be used in contact with food preventing contamination and preserving its organoleptic properties. As a proof-of-principle application, the edible conductive pastes are demonstrated to be effective edible contacts for food impedance analysis, to be integrated for example in smart fruit labels for ripening monitoring.

1. **Introduction**

Myriad of applications related to drug delivery, food tagging, and human health monitoring would benefit from a class of electronic devices that are safely edible.[1-3] In this context, ingestible electronics (IE) represent a powerful tool in preventing, screening, and treating diseases via innovative swallowable devices.[4, 5] As such, IE delivered breakthroughs from the 1950s till today, such as pH-endoradiosonde,[6, 7] ingestible pressure,[8] motion,[9] temperature,[10] and gas sensors,[11] energy sources to power ingestible devices,[12, 13] detectors for the ingestion of capsules or tablets,[14] and cameras[15] and microphones[16] for vital sign monitoring, to cite a few. IE ensures high performance since it uses standard electronics components but simultaneously faces limitations because it is built mainly on materials that are not degradable in the body. Indeed, after ingestion, the non-digestible and bulky nature of standard materials for electronics poses a risk of retention for the patients, requiring medical supervision.[9] Moreover, besides the cost that may force their reuse, the non-degradable, long-lasting, and sometimes poisoning characteristics to the environment impose their recollection after expulsion. Lastly, IE materials and devices employ energy intensive manufacturing that is not sustainable for the single-use desirable for food monitoring purposes.[17]

Edible electronics aim to bridge ingestible with green electronics and shift from swallowable to fully digestible and not hazardous components.[18] Therefore, their implementation will ensure a safe administration without medical supervision and enable point-of-care testing since the device will safely degrade in the body/environment after performing a specific task.[17] Edible electronics envision realizing all the passive and active electronic components exploiting food derivatives, with green and large-scale production methods and enabling sustainable, safe and cheap devices. Nowadays, the most urgent need for edible electronics to thrive depends on creating a library of materials that function as the basic building blocks for eatable technologies. The isolation, the design, and the realization of edible insulators, conductors, and semiconductors that are non-toxic, sustainable, and completely safe for ingestion even in large amounts (i.e., ≥ mg/kg body weight (bw)/day), that are produced through simple and cost-effective processes and exhibit satisfying electrical properties are needed as a first step.

Insulators that are safely edible in large amounts without any adverse impact on human health are plentiful.[17, 19-29] Edible conductors and semiconductors that are not simply ingestible, but are non-toxic, and with good and stable performances are extremely challenging to be developed.[18, 30-32] Studies and technologies entirely dedicated to designing and advancing edible conductors and semiconductors with the features required for edible electronics are rare. Nevertheless, these classes of materials are necessary to enable all the edible passive and active electronic components.

State-of-the-art edible conductors can be categorized into ionic and electronic ones, although mixed conductive behaviors exist. Edible ionic conductors rely mostly on hydrogels. For example, a gel of gellan gum and salted gelatin and commercially available food spread such as Vegemite and Marmite were proposed as 3D printable edible electrodes.[33, 34] Transparent protein-based films of silk/laponite[35] and elastomeric bovine serum albumin[36] were proposed as ionic conductors applied in human-machine interaction and electrophysiological signal sensing, respectively. Melanin thin films and melanin nanoparticles-based composites exhibit a hybrid electronic and ionic conduction mechanism and were suggested as eatable electrodes.[37-41] Yet, edible hydrogels have ionic conductive properties that cannot sustain a constant electronic current in a circuit. Furthermore, the ionic conduction features depend on the environment because relying on hydration.

Typical edible electronic conductors with daily intake in the range of µg/kg bw/day are metals such as magnesium, zinc, iron, silver, or gold. The first three are often used as thin films (i.e., 100 nm - 10 µm) in transient biodegradable electronics applications due to their ability to dissolve at different rates in contact with water-based liquids.[42-45] Magnesium[46] and silver[47] were applied as electrodes for current collection in edible and biodegradable triboelectric nanogenerators. Silver and gold thin films are implemented in miniaturized electronics devices such as honey-gated[18] and transferable on food/pills[17] transistors. Low quantities of gold are also used as current collectors for cheese-[48] and gelatin-[49] based supercapacitors. Nevertheless, metals are pricy and are not compatible with bulk and versatile production methods proper of the polymer industry. According to the European Food Safety Agency (EFSA), they can be eaten safely only in small amounts in the range of 1-100 µg/kg bw/day. Moreover, they are often nanometer sized.[50] Thus, they should be handled with attention in edible applications due to the chance of accumulation in the GI tract that normally increases with decreasing particle size and depending on the particle shape.[1, 50] Another material exploitable as an edible electronic conductor is activated carbon (AC, also known as activated charcoal and vegetable carbon, E 153) that has a daily intake in the order of g/kg bw/day.[1] AC-based electrodes were proposed for edible electrochemical sensors,[51] for sensing glucose in the gastrointestinal (GI) tract,[52] and to realize edible supercapacitors.[48, 49] AC was also proposed as anode and cathode for edible biofuel cells.[53] Generally, AC is readily available on the market at a low price (~ 0.01-0.3 €/g) and is produced on a large scale in powder form, approved by EFSA and Food and Drug Administration. Nevertheless, its powdery form makes it not easy to handle and, as-is, does not present satisfying mechanical and adhesion properties. An edible conductor that is producible on a large scale and at a low cost through green manufacturing methods that are compatible with existing production plants is undocumented. Such edible conductor needs to take advantage from the electronic properties of food and food derivatives that are eatable in large amounts (i.e., ≥ mg/kg bw/day) without any potential adverse impact on human health. It should exhibit stabilities of the electrical properties even in exotic environments such as high humidity environments and water-based ones such as the GI tract. Such electrodes should be compliant and adhesive to different surfaces, including food and pills.

Such an edible conductor would be ideal for food quality monitoring.[54-56] In particular, monitoring the aging of fruits through electrical impedance spectroscopy (EIS) is a potentially disruptive technology to improve the agricultural sector efficiency and reduce food waste.[57] EIS on fruit may enable quick, on the ground, quantitative, and portable analysis of fruits ripening and aging. Thus, it has the potential to substantially improve devices proposed so far, which check the aging through smart labels on food packaging that mainly are constituted by time and temperature indicators or gas sensors.[58-61]

Here we present a conductive composite paste made with materials with a highly tolerable upper intake limit ($\geq$ mg/kg bw/day). The matrix is constituted by a biodegradable oleogel of food-grade beeswax and vegetable oil. Conductivity is provided by adding micrometric-sized AC particles. The composite paste is the first designed specifically for edible electronics. This conductor, displaying a resistivity in the order of 100 $\Omega \cdot$cm, is fabricated without employing solvents and at low temperature, not exceeding 100 °C, ensuring green manufacturing. The paste exhibits tunable electromechanical features and adhesion depending on the composition. The different formulations of the composite are potentially compatible with large-scale production processes such as direct ink writing, extrusion, and injection molding and thus are ideal for an industrial scale-up. Moreover, the paste shows antibacterial activity and is hydrophobic, thus avoiding eventual food contamination when used in contact with it and possessing electrical properties robust to air and water-based environment. To provide an example of the many possible practical uses of the proposed edible conductive paste in the future, we have demonstrated that it enables effective edible contacts for food impedance analysis, to be integrated in smart fruit labels for aging monitoring.

## 2. Results and Discussion

The edible conductor was designed using only food-grade materials eatable in large amounts (i.e., ≥ mg/kg bw/day) without toxic adverse effects. In particular, beeswax (E 901, conservative exposure estimate by EFSA 22 mg/kg bw/day),[62] sunflower oil, and activated carbon (E 153, ≥ 500 mg/kg bw/day)[1, 63] were selected (Figure 1a). The adopted AC was micrometric sized (10-100 microns, see Figure S 1) to avoid any eventual toxicity effect of nanomaterials.[50] The wax and the oil were melt-mixed, adding the desired amount of AC afterward. Solvents were avoided, and the temperature reached during the production process was 100 °C, ensuring a green and straightforward methodology. The thermogravimetric analysis demonstrates that the processing temperature leaves the materials intact, as shown in Figure S 2. Indeed, all the samples start to degrade above 200 °C. At the same time, a temperature of 100 °C ensures a complete melting of the wax, as shown in the differential scanning calorimetry analysis presented in Figure S 3, enabling an efficient mix with oil and AC. Lastly, the employed temperature leaves the chemical fingerprint of wax and oil intact after samples fabrication, as shown by the FTIR analysis of Figure S 4, ensuring unaltered edibility and organoleptic properties of the pastes. The material resulting from this procedure is a composite paste that is pliable and can be shaped as required, as shown in Figure 1b and Video S1. The fabrication process can be easily scaled up with uncomplicated industrial appliances since the materials used are produced in large scale and the composite paste is compatible with polymer manufacturing techniques such as direct ink writing (inset Figure 1b). All the data reported below refers to samples produced by melt mixing.

The micromorphology of the wax-oil mixture (weight ratio 1:1, named WaxOil (1:1) from now on) is displayed in Figure 1c. Wax-oil blends are termed oleogels, and the gelation mechanism is connected to the arrangement of chemical species such as wax esters and alkanes into micrometric plate-like wax assemblies, which aggregate to form 3D networks of crystals of wax that entraps liquid oils.[64, 65] Such microcrystal platelets are evident in the beeswax sunflower oil mixture and are responsible of the micrometric size roughness.[66] Adding the AC filler that has a size range of 10-100 µm (Figure S 1), on the one hand, leads to a more uneven texture owing to the incorporation of the filler inside the matrix. On the other, due to the AC intrinsic porosity that absorbs the oil, the wax microcrystals are more evident with AC inclusion, as it is possible to notice comparing Figure 1c and Figure 1d (see also SEM at variable AC content in Figure S 5).

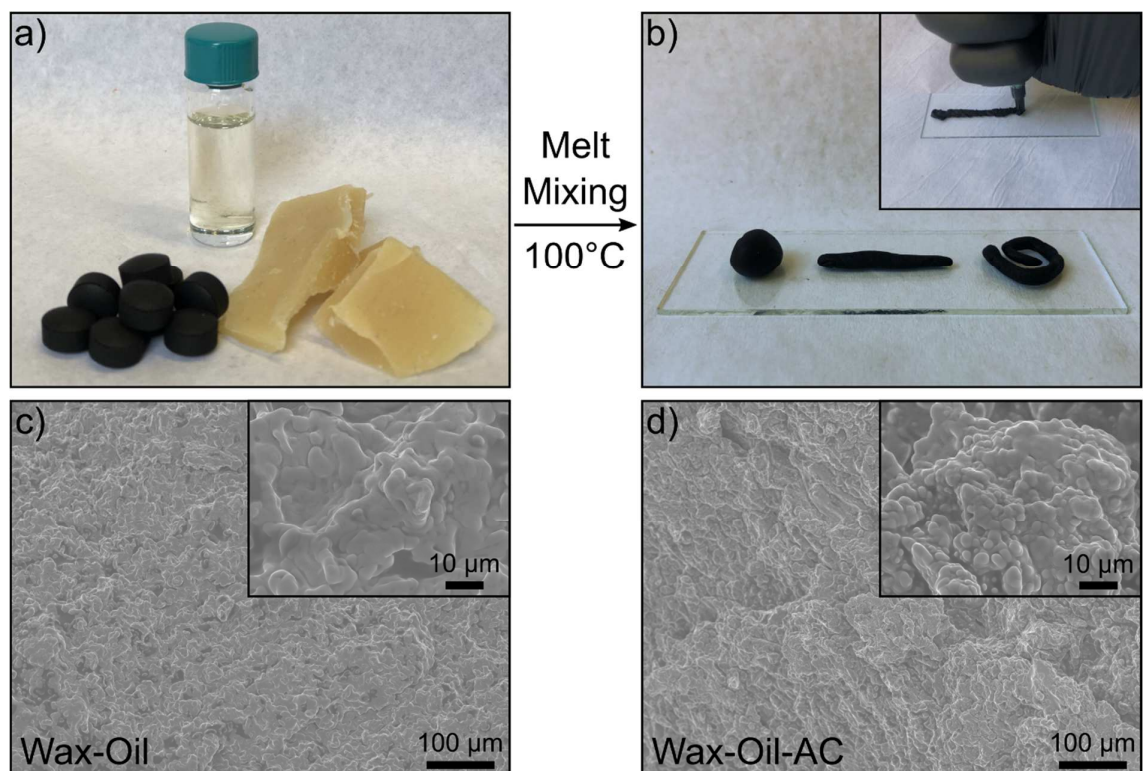

**Figure 1.** Edible conductor composition, manufacturing, and micrometric structure. a) raw materials that constitute the composite paste: Sunflower oil, Beeswax and Activated Carbon (AC). b) Pliability of the edible conductor and demonstration of its potential for direct ink writing techniques (inset). c) - d) Scanning elecron microscopy of the pure wax-oil matrix at 1:1 ratio and edible conductor made adding 30 wt% of AC to the matrix, respectively.

The ability to adapt and tune the rheology of composite pastes through variation of matrix and particle compositions can deliver materials that can be produced with many manufacturing techniques, which display multiple functionalities and are easily adaptable to a specific application.[67-69] Therefore, the rheological properties of the WaxOil (1:1) and the oleogels realized by adding different amounts of AC (from 0 to 40 wt.%), were tested. The samples with AC were labeled simply referring to the wt.% (% from now on) of the AC added to the wax-oil matrix.

Frequency sweep analyses of several rheological parameters (Figure 2 a-d and Figure S 6) were performed between 1 and 100 Hz at three different temperatures: ambient temperature (25°C), 37°C to simulate proximity with the human body and 45°C to simulate a hot natural environment (e.g., a very warm summer). All the oleogels including AC show a solid gel-like viscoelastic behavior with storage modulus G' > G" (loss modulus), G' increasing similarly to G" with increasing frequency, and a phase shift consistently below 45°, as noticeable comparing Figure 2a, Figure 2b, Figure S 6, and Table S 1. The only sample characterized by a fluid state is the pure WaxOil (1:1) oleogel at 45 °C, which displays a phase shift of ~ 51 °, as appreciable in Table S 1. Interestingly, the G' and G" tuning is achieved by changing the percent of AC filler inside the oil wax matrix. Indeed the WaxOil (1:1) oleogel at 25 °C shows a storage modulus of ~ 18 MPa at 1 Hz, gradually reaching a value of ~ 45 MPa by increasing the filler loading to 40 % AC, very close to the one of pure beeswax at ~ 47 MPa, as appreciable in Figure 2a. A similar tuneability but at lower G' values is observed for the samples at 37 °C. In this case, the WaxOil (1:1) sample exhibits a storage modulus of ~ 2.6 MPa, while the sample filled with 40 % AC displays a modulus of ~ 11 MPa, as shown in Figure 2b. The same trend is observed for the rheological tests at 45° C but with even lower values of G' and G", with G' in the order of 0.1 MPa at 1 Hz (see Figure S 6). The moduli of the samples are significantly decreased with increasing temperature because of the proximity of the melting temperature (~ 62°C - Figure S 3) to the testing temperature. These moduli are comparable with those of pasta dough that shows G' in the order of 0.1 MPa at similar testing parameters.[70, 71]

The complex viscosity ($\eta$) of all the samples shows a shear-thinning behavior, as shown in Figure S 8, a feature particularly relevant for pastes, which need to be soft and easily pliable when used or processed, while they need to show high viscosity at rest after application to prevent spreading.[72] $\eta$ can be fine-tuned with respect to the wt.% of AC, and in general it decreases with increasing frequency, as shown in Figure 2c and Figure 2d. At 1 Hz, it increases from ~ $3.0 \cdot 10^6$ to $7.3 \cdot 10^6$ Pa·s at 25 °C and from $4.9 \cdot 10^5$ to $1.8 \cdot 10^6$ Pa·s at 37 °C, going from the WaxOil (1:1) sample to the 40 % AC one.

A critical problem of the composites loaded with a high amount of filler is the difficulty in the mixing and fabrication; in this case, the pronouncedly increased viscosity may result in brittle and not self-standing materials after fabrication. The 40 % AC sample presents a "paste-powder" form and the abovementioned issues. The brittleness of the composite is confirmed by SEM images, which display a texture with multiple micrometric cracks, as shown in Figure S 7. The tuning of the rheological properties for this sample is achieved by changing the wax and oil ratio, which improves the paste melt-miscibility and conformability. For example, at 37 °C (Figure 2e), changing the wax:oil ratio to 1:3 and 1:9 is effective in decreasing the complex viscosity of one and two orders of magnitude, respectively. A similar dependence of the viscosity on the wax:oil ratio is observed at 25 °C and 45 °C (Figure S 6). Accordingly, as shown in Figure S 7, the cracks are absent for the 40% AC-loaded oleogels with 1:3 and 1:9 wax-oil ratios. As such, the self-standing nature of the paste with 40% AC is restored.

Another desirable feature of pastes is the adhesion to various surfaces. Therefore, adhesion on a standard plastic substrate (polymethyl methacrylate, PMMA) and glass is tested for the pure wax, the WaxOil (1:1) and the samples with diverse AC content, as shown in Figure 2f (PMMA substrate) and Figure S 9 (glass substrate). In particular, shear adhesive strength tests demonstrate that the adhesion strength of the wax-oil matrix could be tuned by the inclusion of different amounts of AC inside the composite formulation. Indeed, the pure oleogel displays an adhesive strength on PMMA of ~3.3 kPa and adding AC from 8 till 30 % enhances the paste adhesion from ~3.7 to ~12.4 kPa, a value close to pure wax that reached ~16.4 kPa. The tuneability of the adhesive strength with AC inclusion is achieved thanks to the porosity of the AC filler that "dries" the oil component of the paste, enhancing the available "free" wax fraction in the paste that leads to an adherence improvement, maintaining a self-standing structure. At 40 % AC concentration, the "dry phase" (which consists of the AC and the wax) becomes predominant, and the sample powdery nature negatively influences its adhesive properties (Figure S 7). The sample with 40 % AC inclusion does not display any adhesive properties, but by increasing the oil component in the matrix, and adhesion strength in the order of kPa can be re-established, because the complete absorption of the wax-oil phase by the AC is prevented. In particular, tuning the wax to oil ratio to 1:3 and 1:9 restores an adhesion strength for the sample containing 40% AC of ~3.0 and ~1.4 kPa, respectively. The same behavior is measured on glass substrates, but with slightly lower adhesive strength (Figure S 9 and Video S1). Changing the ratio of wax and oil permits the restoration of adhesive properties fundamental for practical applications. The measured adhesive strength values are comparable, for example, with polymer-based degradable adhesives used for surgery and tissue engineering (range 1-100 kPa)[73, 74] and polyurethane-based adhesive for wood (~ 80 kPa).[75]

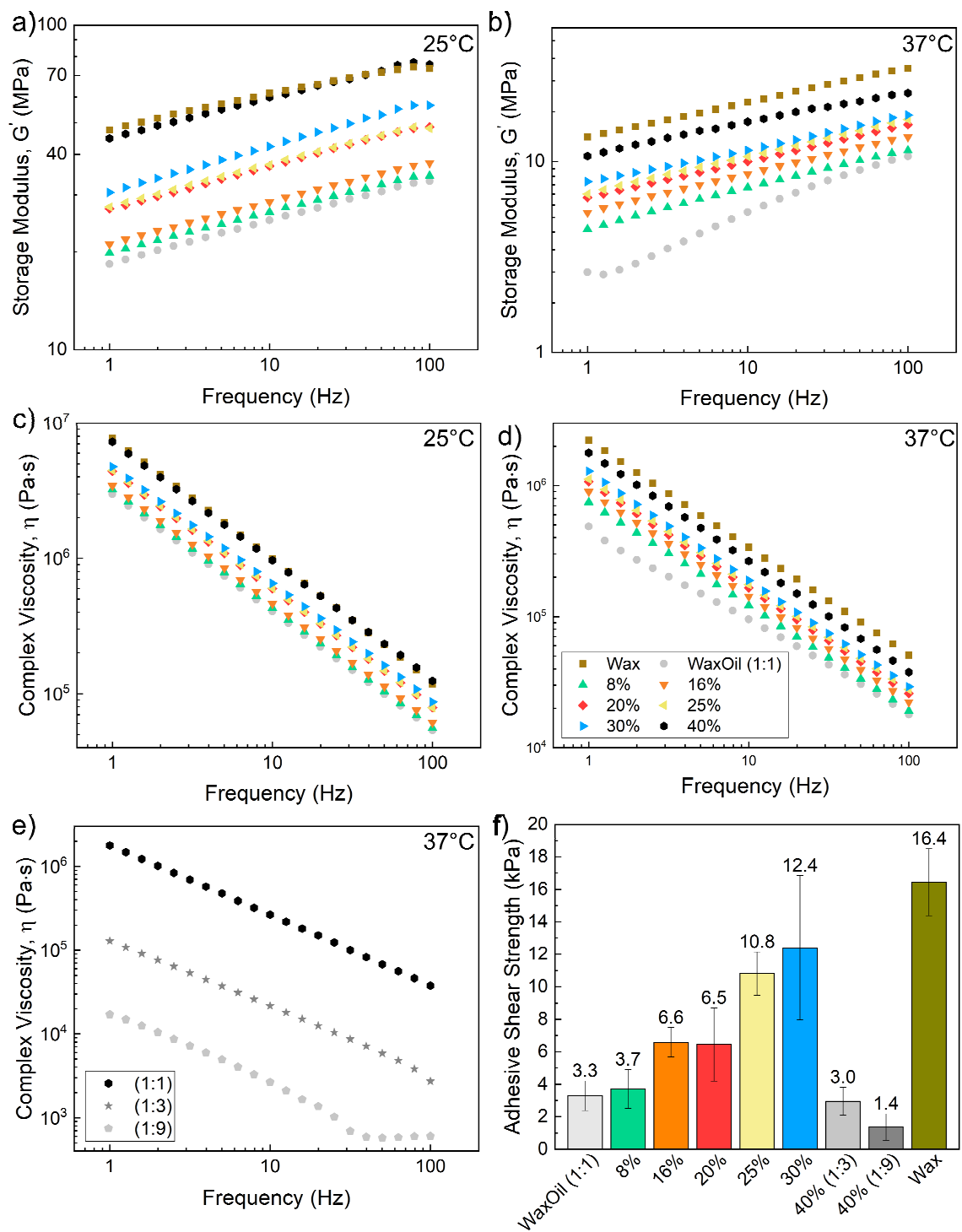

**Figure 2.** Rheological and adhesion features of the composite paste. In Figures a)-d) and f) the pure beeswax (Wax), the wax vegetable oil matrix at a 1:1 ratio (WaxOil (1:1)), and the samples realized adding different amounts of activated carbon (AC, from 0 to 40 wt.%) were tested. a) and b) show the samples' storage moduli (G') at 25 and 37°C, respectively. c) and d) display the complex viscosity η at 25 and 37°C, respectively. e) Tunable η of the composite paste loaded with 40% AC and diverse wax:oil ratios (1:1, 1:3, 1:9). f) Adhesive shear strength of the sample as a function of the composition.

The inclusion of conductive fillers inside insulating matrixes is effective towards inducing electrical conductivity in the resulting composites.[76] The minimum filler loading required to switch the composite from an insulator to a conductor is the electrical percolation threshold.[77-79] The current voltage (I-V) curves of the edible pastes are ohmic from 20% AC concentrations, as shown in Figure S 10, and the majority charge carriers are determined to be holes given the positive measured Seebeck coefficients, typical of graphitic materials,[80] as shown in Table S 2. The edible paste resistance is measured against the weight of AC filler incorporated inside the wax-oil matrix and converted to resistivity following the formula reported in the methods part. At filler loadings up to 16 %, the direct current resistivity of the oleogel is above $10^9$ $\Omega\cdot$cm, corresponding to the resolution limit of the adopted instrument. At 20 % AC, electrical percolation is reached and the resistivity dropped to ~ 1 M$\Omega\cdot$cm. Increasing the filler content to 40% permits to reach values of 101 ± 2 $\Omega\cdot$cm, as shown in Figure 3a. The micrometric fillers inside the matrix at low loadings (i.e., lower than 20% AC) are not densely packed and thus do not create an electrical conduction network. On the other hand, after reaching the electrical percolation threshold, an electrically conductive path is available for the charge carrier, as schematically drawn in the bottom of Figure 3a. All the samples with 40% AC filler display equal resistivity values independently on the ratio between the wax and the oil in the matrix (see Figure S 11), indicating that similar dispersions of the fillers are obtained for all of them after fabrication. The obtained resistivity values are comparable with other reports that employ solely activated carbon as an edible electrode (range of resistivity 0.5-1000 $\Omega\cdot$cm),[1] here with the added advantages of a paste such as pliability, conformability, and compatibility with large scale production methods proper of the composite industry. The conductivity of the 40% AC loaded sample is sufficiently low to enable the powering of a light-emitting diode (LED) electrically connected through the paste, as shown in the inset of Figure 3a and Video S1.

For practical applications, a desirable property for an edible conductor is the stability of the electrical conduction properties even in contact with water-based liquids.[81] Thus, the pastes were immersed in water for more than 50 hours and their weight and electrical resistance variation were monitored. Throughout this period, the weight variation is negligible and the electrical resistance increased by less than 6% of the initial value, as shown in Figure S 12. These two factors reveal a weak interaction of the samples with water, in agreement with the water contact angles on pastes above 90° (see Figure S 13). Such features improve the stability to water immersion of the state-of-the-art that relies on water-soluble metallic particles[1, 81].

The electrical impedance modulus ($|Z|$) and the phase of the edible pastes different formulations in the frequency range $10^2$-$10^6$ Hz are presented in Figure 3b and in Figure 3c, respectively. The pastes with filler loadings below 20 wt. % AC behave as insulators, exhibiting the same curve of an open circuit, therefore implying $|Z| > 10^8$ Ω at a frequency of 100 Hz. Starting from 20 % AC, the paste starts to display a resistive behavior: from 25 % AC, $|Z|$ is constant and in the order of 100 kΩ (up to 5·$10^3$ Hz, after which the instrumental response dominates), with a phase close to zero, as shown in Figure 3c. $|Z|$ further decreases by increasing the AC content maintaining a phase close to zero, showing two orders of magnitude reduction with increasing filler loading up to 40 %. These measurements confirm the quasi-static resistance measurements(Figure 3a): the electrical percolation threshold for this conductive oleogel is close to 20 % AC, and pastes have purely resistive electrical properties after percolation.

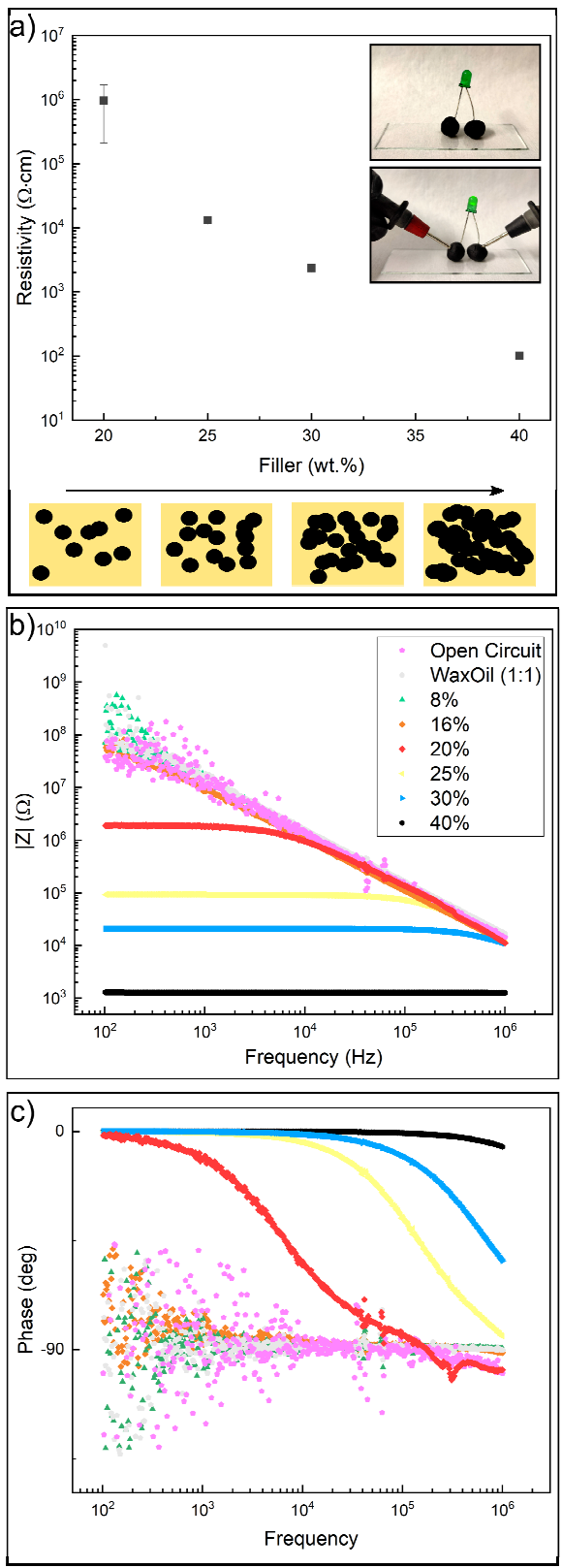

**Figure 3.** Resistivity and impedance of the edible composites. a) Electrical resistivity as a function of the AC content inside the wax-oil oleogel matrix. In the inset, the powering of a green LED through the eatable paste, with 12 V applied. Below, the schematic of the percolation

of a filler inside a matrix. b) Modulus of the sample impedance with different amounts of activated carbon included inside the wax-oil paste. The pure matrix (WaxOil (1:1)) and the open circuit measurements are reported as references. c) Phase of the impedance of the sample reported in b).

As a proof-of-principle of one of the many possible applications of the proposed conductive paste, we have tested it as an electrode in a label made of edible materials to monitor fruit aging through electrical impedance spectroscopy (EIS).[82-86] Indeed, a crucial components to realize an enabling technology for food monitoring based on EIS are the electrodes. The reported studies exploit Ag/AgCl electrodes or other metallic ones.[57, 85, 87] An essential operational part of these electrodes is the electrolyte gels applied between the electrodes and fruits.[85] The hydrogel improves the adhesion and contains free chloride ions such that the charge can be carried to the electrodes. Although these devices are a standard in electrochemistry and clinical practice (e.g., electrocardiogram, electroencephalogram, electromyography), they show drawbacks if applied to food chain supply monitoring. Indeed, ideally, electrodes for EIS on food require two main features:

- edibility for safe implementation: the tag is attached directly on fruit and should be removed before eating the monitored food. As such, it could release some residue on the food. Such residue should be safely edible without any possible adverse effects.
- low cost to provide a widespread usage.

The Ag/AgCl electrodes own none of these characteristics. Indeed, Ag is a noble metal (~0.6 €/g), and its acceptable daily intake is low (µg/kg bw/day),[1] making Ag/AgCl electrodes not suitable for this technology. Electrodes with satisfying compliance and adhesion to different irregular and curved surfaces produced with green and industrially compatible manufacturing processes would facilitate a fast implementation of EIS for food monitoring.

The formulation with 40 % AC and a ratio of wax to oil of 1:3 was selected to fabricate the edible electrodes since it shows the best compromise in rheological properties, adhesion, and electrical conductivity. Apples were selected as model fruits because they are among the most common ones tested so far with EIS.[82-86] The fabrication steps for the edible label comprise the sticking of two conductive electrodes on the fruit, their wiring, and the encapsulation with pure beeswax, as shown in Figure 4a-c. The electrodes are compliant on the apple surface, as shown in Figure 4d. The obtained tag with an area of 1 cm$^2$ thoroughly adheres to the fruit to support the weight of the apple itself (~ 100 g) without detaching (see Figure S 14).

The modulus ($|Z|$) and the phase of the impedance of the apple measured in the frequency range between $10^2$-$10^6$ Hz through the edible electrodes are presented in Figure 4e and Figure 4f, respectively. In 14 days, the impedance increases for the edible composite. Such a trend is already found in previous reports on EIS on fruits.[57, 86, 88] Noteworthy, the EIS performed with edible conductor shows a predominant capacitive behavior (phase ~ 80°) unvaried during the 14 days (Figure 4f). At a fixed frequency of 100 kHz (Figure 4g), the label made with the edible paste gives a value of $|Z|$ of the apple of ~ 18 kΩ at the starting time and reaches a value of ~ 30 kΩ after 14 days. A similar starting value is recorded with a silver paint electrode painted on the same apple and measured simultaneously, reaching ~ 23 kΩ after 14 days.

In Table 2 the list of ingredients of the edible label is reported. Noteworthy, all the components are below the limit of daily intake suggested by EFSA. Therefore, after removal, the toxicity risk is negligible even if there is a bulky residue of the label still attached on food.

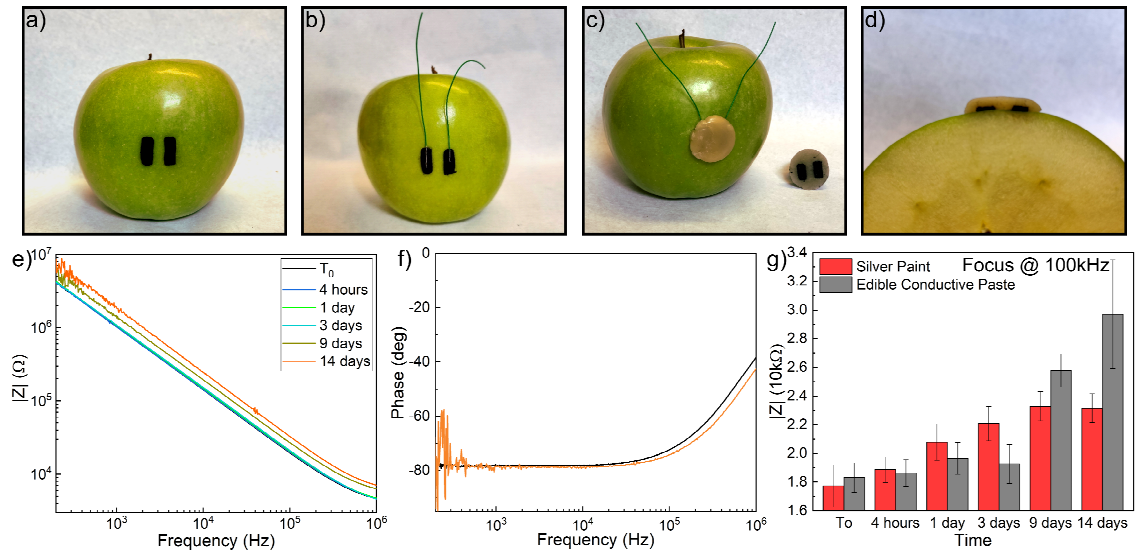

**Figure 4.** The fabrication process of a proof of concept label for electrical impedance spectroscopy (EIS) on fruits and its performances. a)-c) Photos of the steps needed to assemble the sensors on an apple and d) of its cross-section. e) shows the modulus of the impedance of the apples recorded for 14 days through EIS made with the edible tags on the fruit. f) displays the phase of the impedance of the apples recorded through EIS. The phase is constant for 14 days. g) a focus of the time dependent impedance of the apple at 100 kHz recorded with the edible tag (gray column) and using silver paint (red column) as a comparison.

An additional benefit of using beeswax is that it may provide natural antibacterial activity,[89] a property that avoids food contamination. Escherichia coli represents one of the most widespread food contaminants.[90] The results of antimicrobial tests performed on the Gram-negative bacterium *E. coli* are shown in Table 1, reporting the bacterial survival rate for the wax vegetable oil matrix at a 1:1 ratio and the samples realized adding different amounts of activated carbon (8%, 30% and 40 %). After 24 h of incubation time, the uncoated petri dish polystyrene plastic allows the growth of *E. coli* strain until an increment in colony forming unit of $8.44 \cdot 10^8$ CFU/mL. A moderate biocidal effect against *E. coli* is observed when the wax vegetable oil matrix is used. After 24 hours, a low decrement in the cell density is observed on petri dish, confirming moderate antibacterial activity (*P*) of *P*<2 Log (see equation 1 in the methods section for details on the definition of *P*). In contrast, the results demonstrate a remarkable bactericidal effect in the presence of the samples containing activated carbon with a bacterial survival of about $2-4 \cdot 10^6$ CFU/mL, corresponding to an antibacterial activity $\geq 2$ Log. The better antibacterial response of the composite material gives evidence of the biocidal activity that is mainly driven by the activated carbon contribution.

**Table 1.** Logarithm of the number of viable bacteria recovered after 24 hours and antimicrobial activity $P$ against E. coli calculated by Equation (1). Data are reported as the mean ± standard deviation.

| Sample | Log CFU/mL | Antibacterial activity ($P$) |
| --- | --- | --- |
| Uncoated plastic | 8.42 ± 0.05 | |
| WaxOil (1:1) | 7.00 ± 0.04 | 1.42 ± 0.06 |
| AC 8% | 6.20 ± 0.15 | 2.22 ± 0.14 |
| AC 30% | 6.42 ± 0.13 | 2.00 ± 0.12 |
| AC 40% | 6.40 ± 0.03 | 2.02 ± 0.05 |

3. **Conclusion**

Edible electrical conductive pastes made with food-grade materials have been successfully demonstrated. The edible conductors advance the state-of-the-art of edible electronics by being eatable in large amounts ($\geq$ mg/kg bw/day). Furthermore, they are produced with green methods that do not involve solvents and display low energy consumption, i.e. temperatures up to 100°C for a short amount of time, while they are compatible with large-scale production processes. The materials are made with pure food and additives that are already employed in large-scales in the food industry and pharmacology.

An oleogel matrix made with biodegradable and digestible ingredients such as natural waxes and oils and a micrometric eatable conductive filler such as activated carbon constitutes the conductors. These composites can reach a resistivity as low as 100 Ω·cm. Such resistivity does not change upon air and water contact. They exhibit tunable electromechanical features and adhesion depending on the composition. In particular, the activated carbon concentration and the wax to oil ratio ensure precise control of the paste rheological, adhesive, and electrical properties. In addition, they are conformable, pliable, and can adhere to curved surfaces. They feature antibacterial and hydrophobic properties, avoiding food corruption and possible adverse effect upon ingestion and the variation of their properties upon the interaction with the body. The edible conductors are assembled as a proof-of-concept food label that can monitor fruits aging through simple electrical impedance measurements. Such green and sustainable technology could constitute the first building block towards large-scale eatable conductors and monitoring the fruit production chain from the farm to fork, as well as finding wide applications in the pharmacological and healthcare sector.

4. **Experimental Section/Methods**

*Materials*: Refined beeswax was bought from Sigma Aldrich (E 901). Activated carbon was purchased by the same supplier (Supelco, puriss. P.a., powder, E 153). Sunflower Oil and apples were acquired in a local supermarket. These materials were used as received without any further process needed.

To prepare the samples, the wax and the oil were first melt mixed at 100 °C for 20-30 minutes. Afterwards, the required amount of AC was added and mixed at the same temperature and time. The obtained samples were pastes that can be shaped on demand and were proven to be direct ink writable.

To manufacture the proof of principle edible label on fruits, two conductive electrodes were stuck on the apple, wired, and then encapsulated with pure beeswax. The encapsulating wax layer acts as protection and enhance the tag adhesion on the fruits (see Figure 4 and Figure S 14). The electrodes were fully compliant on different surfaces. The amount of the material used for such application was weighted and is summarized in Table 2. Note that if we consider the average weight of an adult person to be 70 kg, the entire tag could be safety eaten.

**Table 2.** Amount of the materials used for the tag on the fruit. The composition used was with 40% AC and a wax:oil ratio of 1:3.

| Sample | Wax (g) | Oil (g) | Activated Carbon (g) |
| --- | --- | --- | --- |
| electrodes | 0.012 ± 0.003 | 0.035 ± 0.006 | 0.019 ± 0.003 |
| Wax | 0.868 ± 0.093 | N.A. | N.A. |

*Methods*: Thermogravimetric analysis (TGA) was performed using a TGA Q500 (TA Instruments) equipped with platinum pans. Each sample was weighted and heated up to 600°C at 10°C/min in nitrogen atmosphere.

Differential Scanning Calorimetry (DSC) was performed using a DSC 25 (TA Instruments) previously calibrated with indium. The samples (20.0±1 mg) were hermetically closed in aluminum pans and underwent a series of heating and cooling steps. Each sample was heated up to 170°C at 10°C/min, held at 170°C for 2 min, cooled down to 30°C at 10°C/min, held at 30°C for 2 min, then heated up again to 170°C at 10°C/min. An empty aluminum pan was used as reference.

Rheology experiments were performed using a Discovery HR-3 Rheometer (TA Instruments) equipped with a Peltier plate temperature system. An aluminium parallel plate geometry with a diameter of 20 mm was used as upper geometry and a gap width of 1.5 mm was maintained for each sample. Samples were prepared on the Peltier plate at 45C to allow optimal shaping and trimmed prior to analysis. Oscillatory amplitude sweep and frequency sweep tests were performed on each sample at 3 different temperatures: 25, 37, and 45 °C. First, amplitude sweep tests were performed to determine the linear viscoelastic region in which the storage modulus (G') and the loss modulus (G") are independent of the strain amplitude. The tests were conducted at a constant frequency of 1 Hz, while varying the torque from 10 to 5000 µN.m (which was translated in oscillation strain % for better understanding of the results). A non-Newtonian behavior was determined with a steep decrease of the moduli beyond the linear viscoelastic region. Secondly, frequency sweep measurements were conducted at a constant torque of 1000 µN.m while varying the frequency from 1 to 100 Hz (which was translated in oscillation frequency for better understanding of the results). Each measurement was performed after 3 min of soaking at the desired temperature to allow the sample to stabilize. Before the beginning of each run, a constant force of 2 N was applied to the sample and maintained throughout the test in order to keep the sample in place and avoid slipping. The only exception was made for high-load-AC samples, which required a 10 N axial force to avoid slipping during measurements at 25°C.

Infrared spectra of the samples were collected through an attenuated total reflectance (ATR) device (Bruker-VERTEX 70v FT-IR Spectrometer). All measurements were recorded with a resolution of 4 cm$^{-1}$, and collecting 128 scans. During measurements, the samples were placed on the spot of the ATR device and gently pressed.

SEM measurements were performed using a Jeol JSM-6010LV operating at an acceleration voltage of 10 kV and recording the secondary electron emission. The distance between the samples and the SEM aperture was 0.5 cm. After immersion at −195.79 °C in liquid nitrogen, the samples were cut by tearing them with two tweezers to preserve the original morphology.

The measurement of adhesive shear strengths was followed with ASTM D3163, regarding bond adhesive of rigid plastic. The adhesive samples studied in this work were stuck onto surfaces of both PMMA and glass slides. The speed of the crosshead was set to be 1.3 mm/min for all samples as suggested by the standard. The values of force were recorded throughout the measurement and then converted into strength with an area of adhesion of approximately 200 mm$^2$ for all the samples tested.

The electrical resistance ($R$) of the pastes was measured using a source-meter from Keithley (model 2450) in the four-probe arrangement. Silver paste (RS pro, product number 186-3600) was painted, forming 10 mm wide contacts ($W$) spaced by 10 mm ($L$) on rectangular shaped specimens 1 mm thick ($t$) sticking on glass slides. Resistivity ($\rho$) was then calculated considering the described geometry trough the formula $\rho = R \frac{W \cdot t}{L}$.

The same setup was used for the measurements underwater but wiring the contacts and encapsulating them with protective beeswax.

The electrical impedance of the composites was measured on specimens with the same dimensions and silver contacts described above. All the measurements were undertaken with Precision Impedance analyzer Agilent 4294A. The frequency of alternating current for the measurement was at a range of $10^2$ to $10^6$ Hz. In the manuscript we report the impedance modulus and phase.

The current–voltage (I-V) curves were measured by means of Semiconductor Parameter Analyzer Agilent B1500A, in four-probe arrangement, employing the same sample described above.

The Seebeck coefficient was determined using the custom-built set-up extensively described in our previous report.[91] Samples were prepared spreading a uniform layer of conductive paste on a glass substrate (2 x 1.5 cm) used as a support. To ensure optimal electrical contact two electrodes were patterned above the film using silver paint (RS PRO Conductive Paint) at a distance of 0.5 cm to comply with the set-up geometry. The measurements were performed at room temperature (25°C) and in vacuum ($10^{-3}$ bar) to reduce undesired thermal convection and conduction phenomena.

Static water contact angles of the pastes were measured by an optical contact angle device (DataPhysics). Ten microliters of deionized water were deposited on the samples.

The biocidal effect of the wax vegetable oil matrix at a 1:1 ratio (WaxOil (1:1)) and the samples realized adding different amounts of activated carbon (AC 8%, 30% and 40 %) on microorganism survival was evaluated as described in the ISO 22196. Briefly, *Escherichia coli* was pre-inoculated aerobically for 12 h at 37°C in Luria-Bertani (LB), with constant shaking at 210 rpm. Bacteria were collected by centrifugation for 10 min at 3500 g and re-suspended at cellular density of about $10^5$ colony forming units (CFU/mL). 0,1 mL of bacteria suspension was deposited on a working area of 4 cm$^2$ of control (uncoated plastic) and of treated samples placed into a separated sterile petri dish. The deposited inoculum was covered with a piece of inert film (polycarbonate film) and the samples were incubated at a temperature of 35 ± 1°C and at a relative humidity of 90%. After 24 hours, 10 ml of phosphate-buffered saline (PBS) is added to the petri dish and a serial dilution was performed. To quantify the viable bacteria, the diluted suspensions were plated on LB-agar plates and incubated for 24 h at 37°C. Subsequently, the number of CFU/mL was quantified for each sample and the bacteria survival was determined as the percentage variation over time of CFUs with respect to the time t = 24 h. The obtained CFUs were used to calculate the antibacterial activity (*P*), according to the formula:

$$P = (U_t - U_0) - (A_t - U_0) = U_t - A_t \qquad (1)$$

Where $U_0$ and $U_t$ are the averages of the common logarithm of the number of viable bacteria recovered from the uncoated plastic immediately after inoculation and after 24h, respectively; while $A_t$ is the average of the common logarithm of the number of viable bacteria recovered from the treated samples after 24 h. The antimicrobial activity can be considered good when $P \geq 2$, and excellent when $P \geq 3$.

*Statistical Analysis*: The data with an error bar are presented as mean ± standard deviation calculated on a minimum of three independent samples using the Origin software.

**Acknowledgements**

Pietro Cataldi and Leonardo Lamanna are equally contributing first authors. This work has been supported by the European Research Council (ERC) under the European Union's Horizon 2020 research and innovation programme "ELFO", Grant Agreement No. 864299. This work has also received funding under the European Union's Horizon 2020 research and innovation programme "GREENELIT", Grant Agreement No. 951747. This work falls within the Sustainability Activity of Istituto Italiano di Tecnologia. The preparation of the samples for SEM measurements was carried out at PoliFab, the micro and nano-technology center of the Politecnico di Milano.